\documentclass[twocolumn,showpacs,preprintnumbers,amsmath,amssymb,superscriptadd ress]{revtex4}

\usepackage{natbib}
\usepackage{amssymb}

\usepackage{graphicx}
\usepackage{dcolumn}
\usepackage{bm}

\def\bea{\begin{eqnarray}}
\def\eea{\end{eqnarray}}

\begin{document}


\preprint{Version 2.15}

\title{Hadronization geometry from net-charge angular correlations on momentum subspace $(\eta,\phi)$ in Au-Au collisions at $\sqrt{s_{NN}}$ = 130 GeV}



\affiliation{Argonne National Laboratory, Argonne, Illinois 60439}
\affiliation{University of Bern, 3012 Bern, Switzerland}
\affiliation{University of Birmingham, Birmingham, United Kingdom}
\affiliation{Brookhaven National Laboratory, Upton, New York 11973}
\affiliation{California Institute of Technology, Pasadena, California 91125}
\affiliation{University of California, Berkeley, California 94720}
\affiliation{University of California, Davis, California 95616}
\affiliation{University of California, Los Angeles, California 90095}
\affiliation{Carnegie Mellon University, Pittsburgh, Pennsylvania 15213}
\affiliation{Creighton University, Omaha, Nebraska 68178}
\affiliation{Nuclear Physics Institute AS CR, 250 68 \v{R}e\v{z}/Prague, Czech Republic}
\affiliation{Laboratory for High Energy (JINR), Dubna, Russia}
\affiliation{Particle Physics Laboratory (JINR), Dubna, Russia}
\affiliation{University of Frankfurt, Frankfurt, Germany}
\affiliation{Institute of Physics, Bhubaneswar 751005, India}
\affiliation{Indian Institute of Technology, Mumbai, India}
\affiliation{Indiana University, Bloomington, Indiana 47408}
\affiliation{Institut de Recherches Subatomiques, Strasbourg, France}
\affiliation{University of Jammu, Jammu 180001, India}
\affiliation{Kent State University, Kent, Ohio 44242}
\affiliation{Lawrence Berkeley National Laboratory, Berkeley, California 94720}
\affiliation{Massachusetts Institute of Technology, Cambridge, MA 02139-4307}
\affiliation{Max-Planck-Institut f\"ur Physik, Munich, Germany}
\affiliation{Michigan State University, East Lansing, Michigan 48824}
\affiliation{Moscow Engineering Physics Institute, Moscow Russia}
\affiliation{City College of New York, New York City, New York 10031}
\affiliation{NIKHEF and Utrecht University, Amsterdam, The Netherlands}
\affiliation{Ohio State University, Columbus, Ohio 43210}
\affiliation{Panjab University, Chandigarh 160014, India}
\affiliation{Pennsylvania State University, University Park, Pennsylvania 16802}
\affiliation{Institute of High Energy Physics, Protvino, Russia}
\affiliation{Purdue University, West Lafayette, Indiana 47907}
\affiliation{Pusan National University, Pusan, Republic of Korea}
\affiliation{University of Rajasthan, Jaipur 302004, India}
\affiliation{Rice University, Houston, Texas 77251}
\affiliation{Universidade de Sao Paulo, Sao Paulo, Brazil}
\affiliation{University of Science \& Technology of China, Anhui 230027, China}
\affiliation{Shanghai Institute of Applied Physics, Shanghai 201800, China}
\affiliation{SUBATECH, Nantes, France}
\affiliation{Texas A\&M University, College Station, Texas 77843}
\affiliation{University of Texas, Austin, Texas 78712}
\affiliation{Tsinghua University, Beijing 100084, China}
\affiliation{Valparaiso University, Valparaiso, Indiana 46383}
\affiliation{Variable Energy Cyclotron Centre, Kolkata 700064, India}
\affiliation{Warsaw University of Technology, Warsaw, Poland}
\affiliation{University of Washington, Seattle, Washington 98195}
\affiliation{Wayne State University, Detroit, Michigan 48201}
\affiliation{Institute of Particle Physics, CCNU (HZNU), Wuhan 430079, China}
\affiliation{Yale University, New Haven, Connecticut 06520}
\affiliation{University of Zagreb, Zagreb, HR-10002, Croatia}

\author{J.~Adams}\affiliation{University of Birmingham, Birmingham, United Kingdom}
\author{M.M.~Aggarwal}\affiliation{Panjab University, Chandigarh 160014, India}
\author{Z.~Ahammed}\affiliation{Variable Energy Cyclotron Centre, Kolkata 700064, India}
\author{J.~Amonett}\affiliation{Kent State University, Kent, Ohio 44242}
\author{B.D.~Anderson}\affiliation{Kent State University, Kent, Ohio 44242}
\author{D.~Arkhipkin}\affiliation{Particle Physics Laboratory (JINR), Dubna, Russia}
\author{G.S.~Averichev}\affiliation{Laboratory for High Energy (JINR), Dubna, Russia}
\author{S.K.~Badyal}\affiliation{University of Jammu, Jammu 180001, India}
\author{Y.~Bai}\affiliation{NIKHEF and Utrecht University, Amsterdam, The Netherlands}
\author{J.~Balewski}\affiliation{Indiana University, Bloomington, Indiana 47408}
% ...

\collaboration{STAR Collaboration}\noaffiliation



\begin{abstract}
We present the first measurements of charge-dependent correlations on angular difference variables $\eta_1 - \eta_2$ (pseudorapidity) and $\phi_1 - \phi_2$ (azimuth) for primary charged hadrons with transverse momentum $0.15 \leq p_t \leq 2$~GeV/$c$ and $|\eta| \leq 1.3$ from Au-Au collisions at $\sqrt{s_{NN}} = 130$~GeV. We observe correlation structures not predicted by theory but consistent with evolution of hadron emission geometry with increasing centrality from one-dimensional fragmentation of color strings along the beam direction to an at least two-dimensional hadronization geometry along the beam and azimuth directions of a hadron-opaque bulk medium.
\end{abstract}





\maketitle

\section{Introduction}

Analysis of correlations and fluctuations plays an important role in studies of the colored medium produced in ultrarelativistic heavy-ion collisions~\cite{stock,poly,dcc}. {\em In-medium modification} of parton scattering and fragmentation of energetic partons by the bulk medium produced in heavy-ion collisions may significantly alter large-momentum-scale two-particle correlations relative to those observed in p-p collisions. Large-momentum-scale correlations may result from initial-state multiple scattering~\cite{iss,jetquench}, in-medium dissipation of scattered energetic partons~\cite{newref} and hadronization of the colored medium to final-state hadrons (fragmentation of color strings in p-p, hadronization of the bulk medium in A-A). The local geometry of hadronization, which can be accessed by net-charge correlations, is the subject of this paper. 

String fragmentation models~\cite{lund} describe two-particle correlations on pseudorapidity and azimuth $(\eta,\phi)$ in high-energy p-p collisions in terms of local conservation of transverse momentum and net charge leading to canonical suppression of event-wise net-momentum and net-charge fluctuations. The nature of the corresponding process in A-A collisions remains an open question. Some change should be expected in the correlation structure as the medium evolves from that produced in very peripheral collisions (approximating minimum-bias proton-proton collisions) to that in central heavy-ion collisions.  Predictions have been made of dramatic suppression of net-charge {\em fluctuations} in central A-A collisions as a signal of quark-gluon plasma formation~\cite{jeon}. The question arises what detailed net-charge {\em correlation structure} would correspond to such predictions, and what structure is actually present in heavy-ion collisions.

In this Letter we report the first measurements in heavy-ion collisions of the centrality dependence of two-particle {\em charge-dependent} (net-charge) correlations on angular subspace $(\eta,\phi)$, where charge-dependent here refers to the difference between correlations for like-charge-sign pairs and unlike-sign pairs. 
This analysis is based on Au-Au collisions at $\sqrt{s_{NN}} = 130$~GeV obtained with the STAR detector at the Relativistic Heavy Ion Collider (RHIC). The observed correlation structure suggests that local charge conservation at hadronization combined with increasing system density and spatial extent results in evolution with Au-Au centrality from one-dimensional (1D) {\em charge-ordering} (locally alternating charge signs) on configuration space $z$ (the collision axis), coupled to $p_z$ (or pseudorapidity $\eta$) by longitudinal Bjorken expansion, to two-dimensional (2D) charge ordering on beam and azimuth directions $(z,\phi)$. Those results have not been anticipated by theoretical models~\cite{jetquench,rqmd}. 

\section{Analysis Method}

We wish to access the complete {\em charge-dependent} (CD) structure of two-particle density $\rho(\vec{p_1},\vec{p_2})$ with minimal distortion and without imposition of a correlation model. In this analysis of net-charge {\em angular} correlations we project the two-particle momentum space onto angular subspace $(\eta_1,\eta_2,\phi_1,\phi_2)$ by integrating over a specific transverse momentum interval. The structure of net-charge correlations on {transverse momentum} with specific angular constraints will be considered in a future analysis. 

Correlations are obtained with a {\em differential} analysis which compares object and reference pair density distributions. The object distribution is comprised of particle pairs formed from single events, referred to as {\it sibling} pairs, and the reference distribution consists of pairs combining particles from two different but similar events, referred to as {\it mixed} pairs. The corresponding pair densities are denoted by $\rho_{sib}(\vec{p}_1,\vec{p}_2)$ and $\rho_{mix}(\vec{p}_1,\vec{p}_2)$ respectively.  The two-particle correlation function $C$ (as commonly defined in nuclear physics) and pair-number density ratio $r$ (as used in the study of quantum correlations or HBT~\cite{starhbt}) are then defined and related by
\bea
C(\vec{p}_1,\vec{p}_2) & = &
\rho_{sib}(\vec{p}_1,\vec{p}_2) - \rho_{mix}(\vec{p}_1,\vec{p}_2) \nonumber \\
 & = & \rho_{mix}(\vec{p}_1,\vec{p}_2)\, (r(\vec{p}_1,\vec{p}_2) - 1),
\label{Eq1}
\eea
with $r \equiv  \rho_{sib}/ \rho_{mix}$. Difference $r - 1$ is the correlation measure we use. In order to visualize the CD correlation structure in the 4D angular subspace $(\eta_1,\eta_2,\phi_1,\phi_2)$ pair densities can be projected onto separate 2D subspaces ($\eta_1,\eta_2$) and ($\phi_1,\phi_2$).  Those projections, discussed further below, discard a substantial amount of the information in the full two-particle space. However, they reveal that significant variation is restricted to {\em difference variables} $\eta_\Delta \equiv \eta_1 - \eta_2$ and $\phi_\Delta \equiv \phi_1 - \phi_2$ (the notation is explained in Sec.~\ref{2part}).  For this analysis we therefore {\em simultaneously} project the 4D subspace onto those angular difference variables. The resulting 2D distribution is referred to as a {\em joint  autocorrelation}. An autocorrelation is a projection {\em by averaging}~\cite{average} from subspace $(x_1,x_2)$ onto difference variable $x_\Delta = x_1 - x_2$. A {\em joint} autocorrelation is a simultaneous projection onto two difference variables. The result of this projection technique is a {\em nearly lossless} (distortion free) projection from the initial 4D angular subspace onto a 2D autocorrelation space.

In this analysis, sibling and mixed pair-number densities $\rho(\vec{p}_1,\vec{p}_2)$ for four charge-pair combinations $(++,+-,-+,--)$ were projected onto $(\eta_1,\eta_2)$, $(\phi_1,\phi_2)$ and $(\eta_\Delta,\phi_\Delta)$.  The projection was done by filling histograms of pair numbers $n_{ab} \simeq  \epsilon_x \, \epsilon_y \,\rho(x_a,y_b)$, where subscripts $ab$ denote the 2D bin indices and $\epsilon_x, \epsilon_y$ are histogram bin widths on variables $x,y \in \{\eta_1, \eta_2, \phi_1, \phi_2, \eta_\Delta, \phi_\Delta\}$. Sibling and mixed pair-number histograms for each charge-pair combination were separately normalized to the total number of detected pairs in each centrality class: $\hat n_{ab,sib} = n_{ab,sib} / \sum_{ab} n_{ab,sib}$ and $\hat n_{ab,mix} = n_{ab,mix} / \sum_{ab} n_{ab,mix}$. Normalized pair-number ratios $\hat{r}_{ab} = \hat{n}_{ab,sib}/\hat{n}_{ab,mix}$ are the basis for this analysis. 

To reduce systematic error, ratio histograms were obtained for subsets of events within a given centrality class which have similar multiplicities (differences $\leq 50$) and primary collision vertex locations within the detector (within 7.5~cm along the beam axis). Ratios $\hat r_{ab}$ for each centrality class were defined as weighted (by total number of sibling pairs) averages over all subsets in that centrality class. Ratios were further combined to form like-sign (LS: $++,--$), unlike-sign (US: $+-,-+$), and charge-dependent (CD = LS $-$ US) ratios.  In this analysis we adopt a CD sign convention compatible with standard particle physics isospin convention and net-charge fluctuation measures~\cite{meanptprl}.

\section{Data}

Data for this analysis were obtained with the STAR detector~\cite{star}
using a 0.25~T uniform magnetic field parallel to the beam axis.
A minimum-bias event sample required coincidence of two Zero-Degree Calorimeters (ZDC); a 0-15\% of total cross section event sample was defined by a threshold on the Central Trigger Barrel (CTB), with ZDC coincidence. Event triggering and charged-particle measurements with the Time Projection Chamber (TPC) are described in \cite{star}. Tracking efficiencies, event and track quality cuts and primary-particle definition are described in~\cite{meanptprl,spectra}. Charged particles were accepted in $|\eta| \leq 1.3$, full azimuth and transverse momentum ($p_t$) range $0.15 \leq p_t \leq 2$~GeV/$c$. Particle identification was not implemented but charge sign was determined. 
Corrections were made to ratio $\hat r$ for two-track inefficiencies due to overlapping space points in the TPC (merging) and intersecting trajectories reconstructed as $>2$ particles (splitting) by applying two-track proximity cuts in the TPC to both $\rho_{sib}$ and $\rho_{mix}$ similar to that done in HBT analyses. 

Small-momentum-scale correlation structures due to quantum, Coulomb and strong-interaction correlations~\cite{starhbt} were suppressed by eliminating sibling {\em and} mixed track pairs ($\sim$22\% of total) with $|\eta_{\Delta}| < 1.0$, $|\phi_{\Delta}| < 1.0$ and $|p_{t1} - p_{t2}| < 0.2$~GeV/$c$ if $p_t < 0.8$~GeV/$c$ for either particle. Those cuts do not significantly affect the correlation structures shown here. Four centrality classes for 300k events labeled (a) - (d) for central to peripheral were defined by cuts on TPC track multiplicity $N$ within the acceptance defined here relative to minimum-bias event multiplicity frequency distribution upper half-maximum end-point $N_0$, which corresponds to the maximum participant number~\cite{meanptprl,nu}. Four centrality classes were defined by (d) $0.03 < N/N_0 \leq 0.21$, (c) $0.21 < N/N_0 \leq 0.56$, (b) $0.56 < N/N_0 \leq 0.79$ and (a) $N/N_0 > 0.79$.
\begin{figure}[h]
\includegraphics[keepaspectratio,width=1.65in]{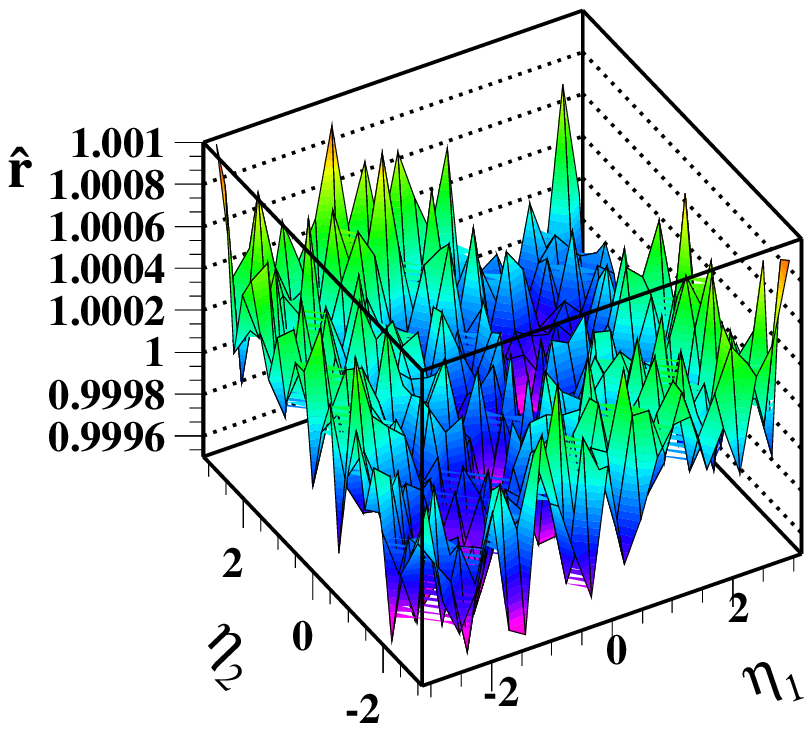}
\includegraphics[keepaspectratio,width=1.65in]{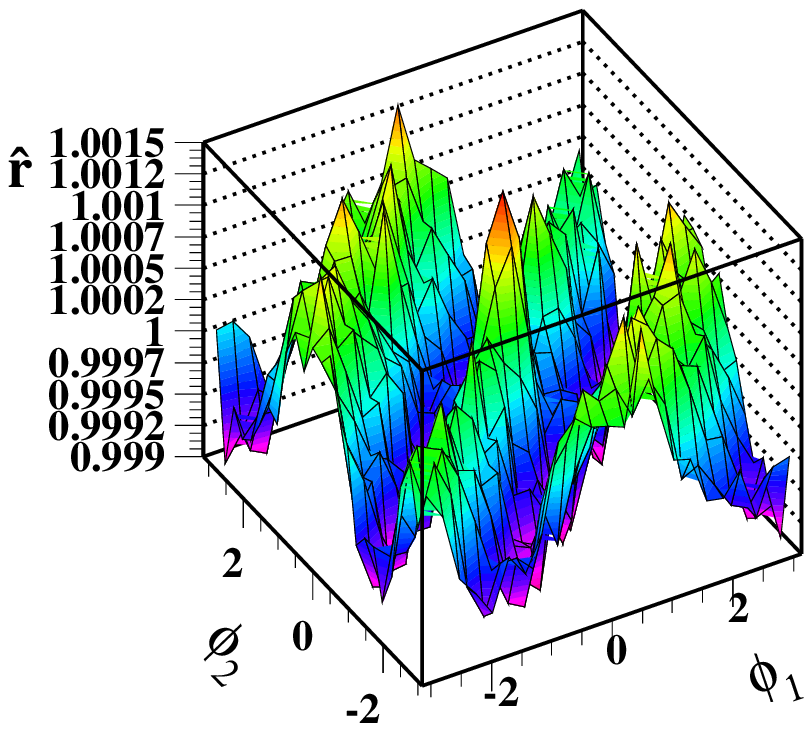}
\caption{\label{Figure1}
Normalized LS pair-number ratios $\hat{r}$ for collisions in centrality class (a) (most central) for $(\eta_1,\eta_2)$ (left panel) and $(\phi_1,\phi_2)$ (right panel).
}
\end{figure}

\section{Two-particle Distributions} \label{2part}

\begin{figure}[t]
\includegraphics[keepaspectratio,width=1.65in]{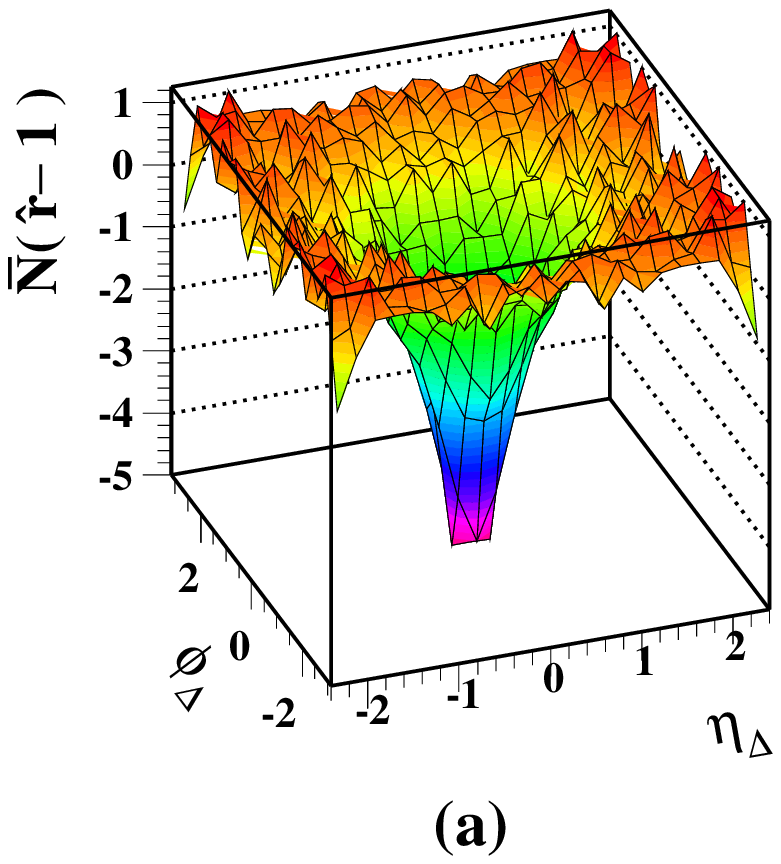}
\includegraphics[keepaspectratio,width=1.65in]{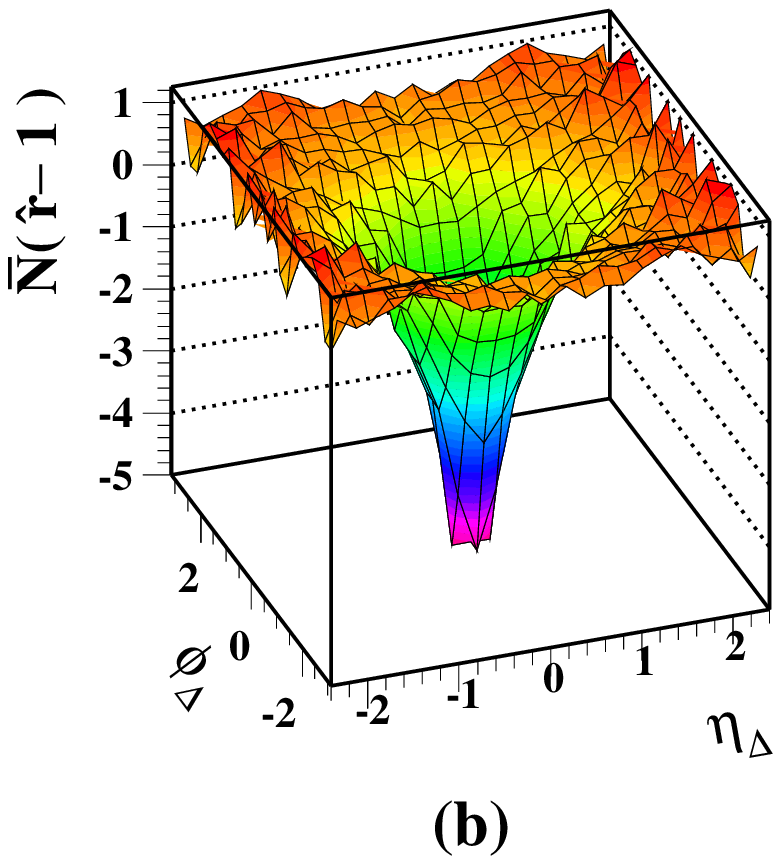}
\includegraphics[keepaspectratio,width=1.65in]{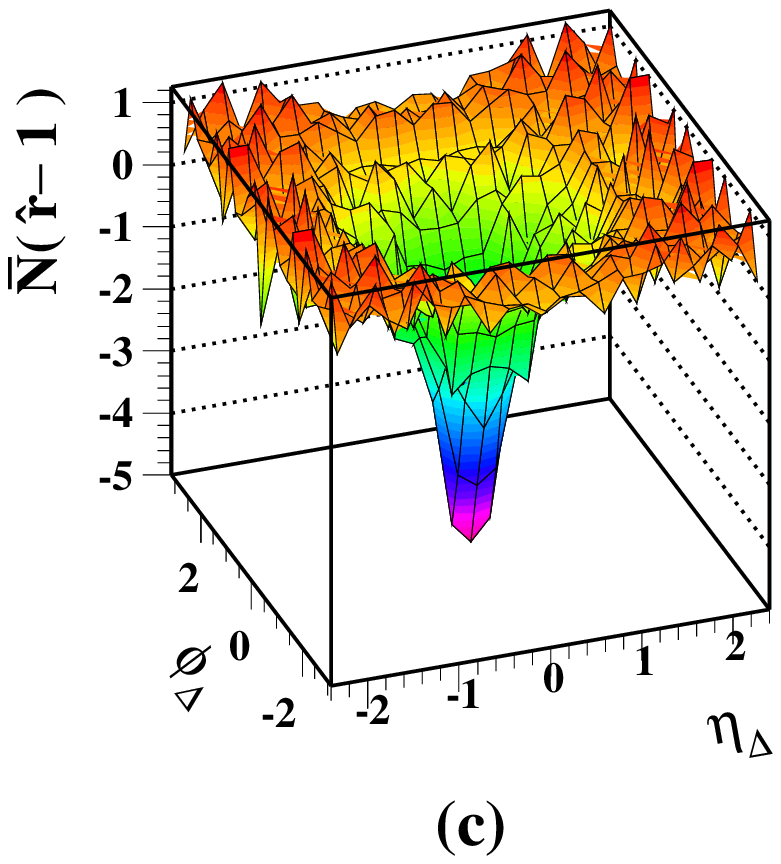}
\includegraphics[keepaspectratio,width=1.65in]{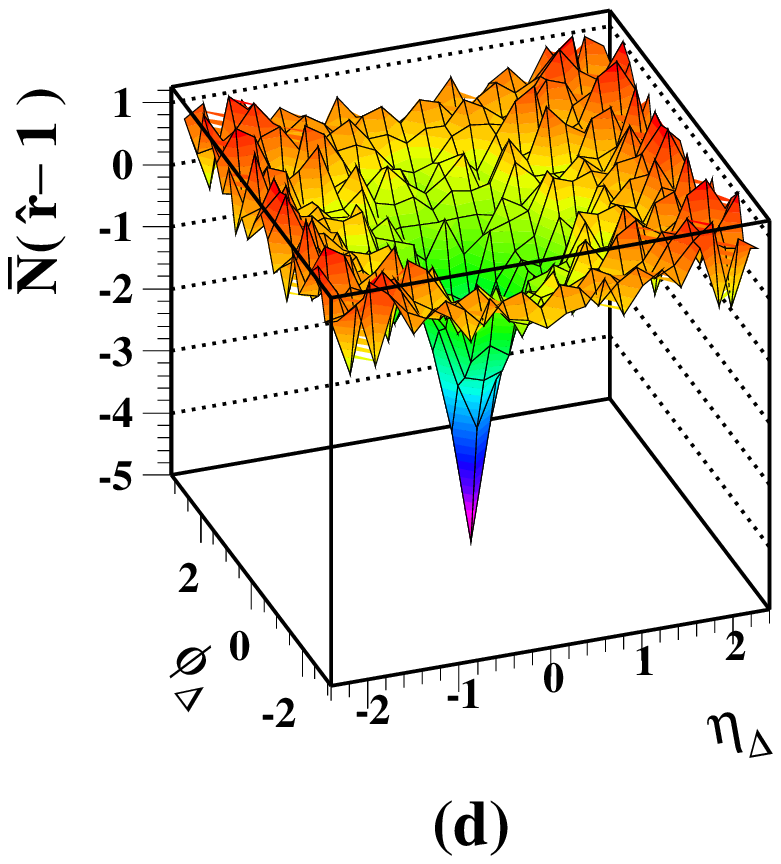}
\caption{\label{Figure2}
Perspective views of two-particle CD joint autocorrelations $\bar N (\hat r-1)$ on  $(\eta_{\Delta},\phi_{\Delta})$ for central (a) to peripheral (d) collisions. Center bins at $\phi_{\Delta} = \eta_{\Delta} = 0$, containing photon-conversion electron pairs, were omitted from model fits.}
\end{figure}

\begin{figure}[t]
\includegraphics[width=1.65in,height=1.65in]{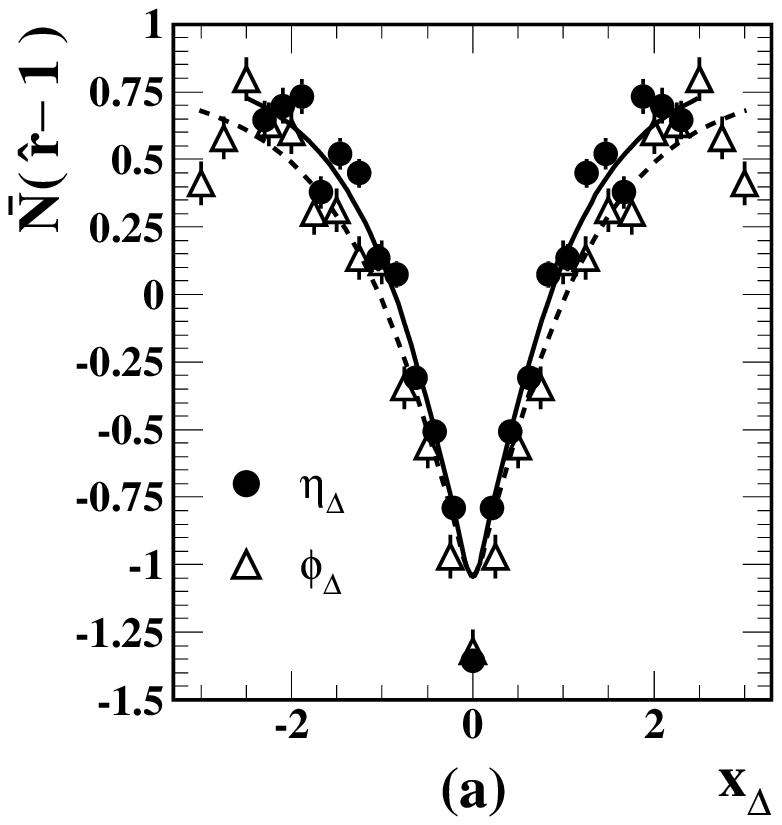}
\includegraphics[width=1.65in,height=1.65in]{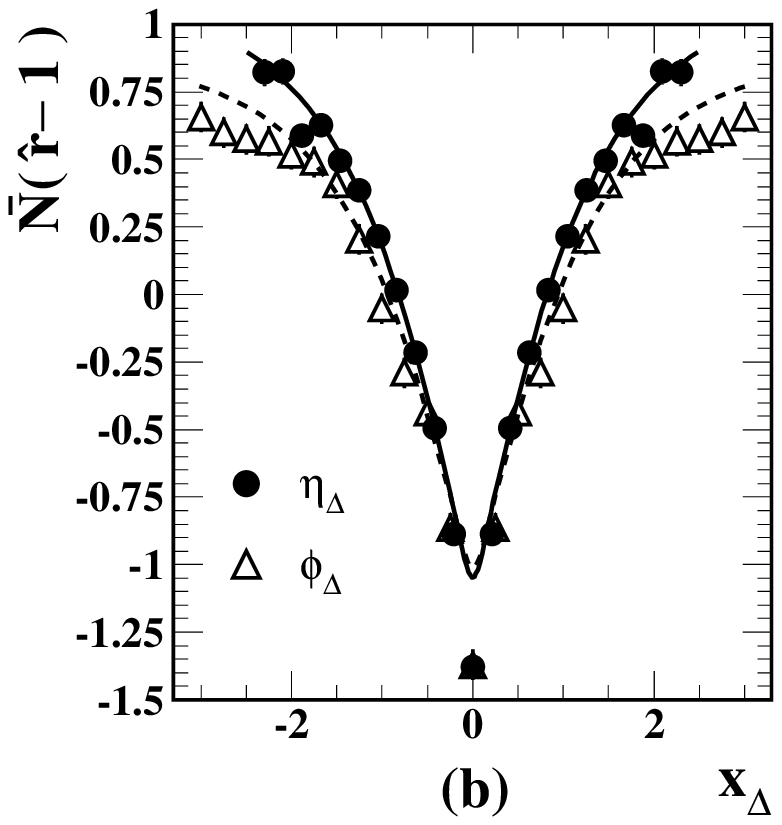}
\includegraphics[width=1.65in,height=1.65in]{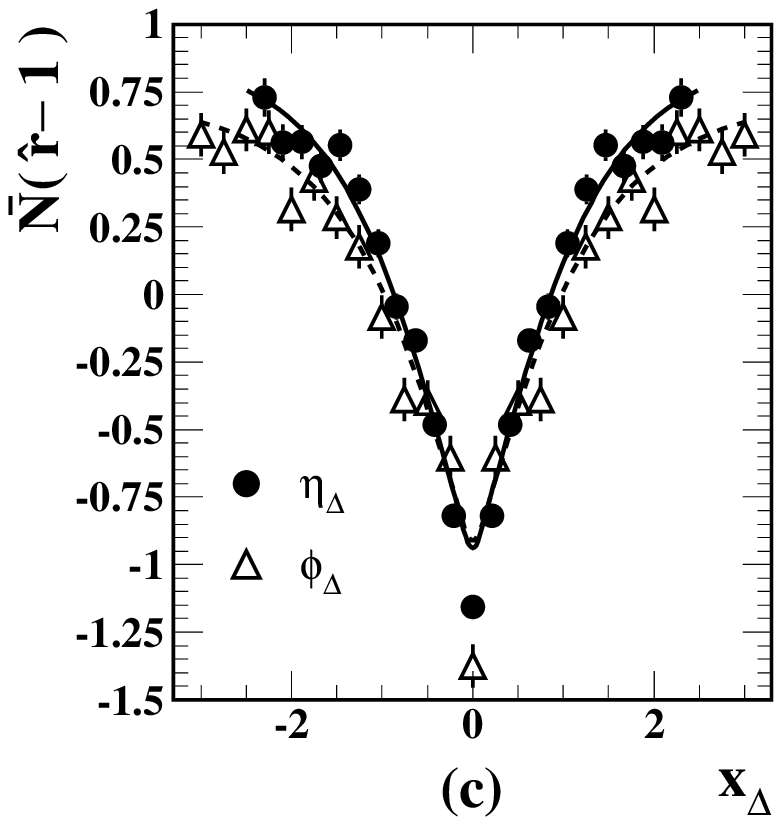}
\includegraphics[width=1.65in,height=1.65in]{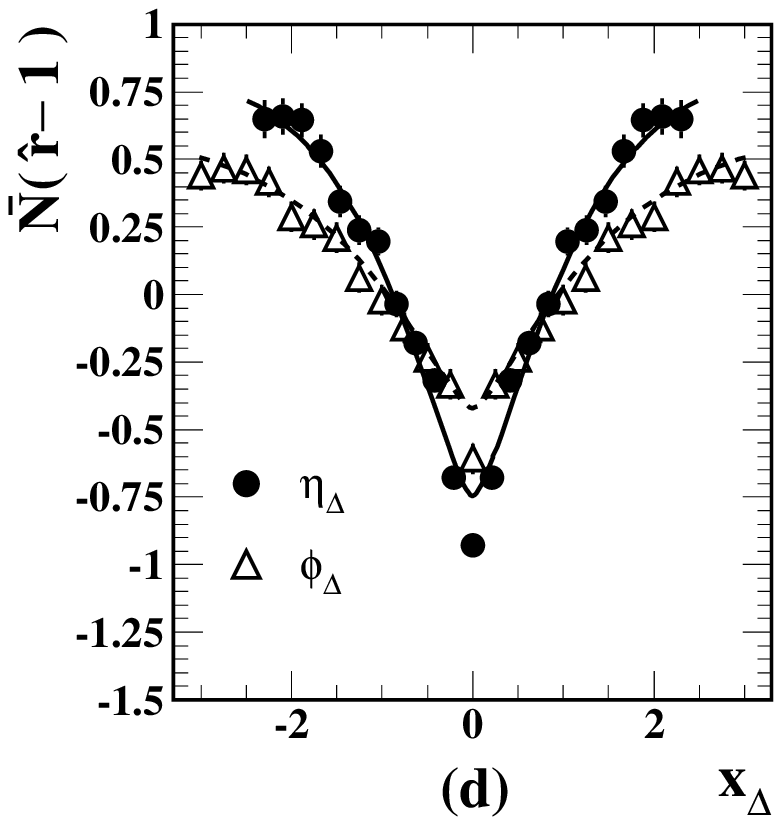}
\caption{\label{Figure3} 
Projections of 2D CD autocorrelations $\bar N(\hat{r} - 1 )$
in Fig.~\ref{Figure2} onto individual difference variables $\eta_{\Delta}$ (solid dots) and $\phi_{\Delta}$ (open triangles) for central (a) to peripheral (d) collisions. Solid (dashed) curves represent projections of 2D analytical model fits to data on $\eta_{\Delta}$ ($\phi_{\Delta}$). The 2D negative peaks are substantially reduced in amplitude after projecting onto 1D.
}
\end{figure}

Fig.~\ref{Figure1} shows ratio histograms $\hat r_{ab}$ for the LS charge combination on $(\eta_1,\eta_2)$ and $(\phi_1,\phi_2)$ for
the most central event class, denoted (a). Deviations from unity ($\hat r - 1$) of this {\em per-pair} correlation measure contain a {\em dilution factor}~\cite{dilution} $1/\bar N$ ($\bar N$ is defined as the mean multiplicity in the detector acceptance) and are therefore numerically a few {\em permil} for central Au-Au collisions. However, the correlation structure is large compared to statistical errors ({\em cf.} Figs.~\ref{Figure2}-\ref{Figure4}). A sinusoid associated with elliptic flow (consistent with conventional reaction-plane measurements) 
dominates the $(\phi_1,\phi_2)$ correlations in the right panel. The {\em anti}\,correlated LS distribution on $(\eta_1,\eta_2)$ in the left panel (anticorrelated: depression along the $\eta_1 = \eta_2$ diagonal) suggests charge ordering from longitudinal string fragmentation as in p-p collisions~\cite{lund,isrpp}.  However, these correlations projected separately onto $(\eta_1,\eta_2)$ and $(\phi_1,\phi_2)$ are incomplete, and quite misleading for A-A collisions. A more complete picture is obtained from 2D {joint} autocorrelations on difference variables $(\eta_\Delta,\phi_\Delta)$ as shown in Fig.~\ref{Figure2}.

Because of the symmetry of these distributions on the angular spaces $(x_1,x_2)$ their description is more natural on {\em diagonal} sum and difference variables $x_\Sigma$ and $x_\Delta$ (reserving conventional difference notation $\Delta x$ for displacement on a 1D space $x$). The invariance of correlation structure on sum variables $\eta_\Sigma \equiv \eta_1 + \eta_2$ and $\phi_\Sigma \equiv \phi_1 + \phi_2$ in Fig.~\ref{Figure1} ({\em i.e.,} parallel to the $\eta_1 = \eta_2$ or $\phi_1 = \phi_2$ diagonals) implies that each distribution can be projected onto its difference variable $\phi_{\Delta} \equiv \phi_1 - \phi_2$ and $\eta_{\Delta} \equiv \eta_1 - \eta_2$ to form an autocorrelation {\em without loss of information}. 
The projection is done by averaging bin contents along each diagonal in Fig.~\ref{Figure1} parallel to the sum axis ({\em e.g.} the $\eta_1 = \eta_2$ diagonal) to obtain the bin contents of a 1D autocorrelation on $\eta_{\Delta}$ or $\phi_{\Delta}$ (the difference axes). Autocorrelation details are described in~\cite{inverse,mitmeths}. If projections are made simultaneously onto both difference variables of Fig.~\ref{Figure1} the resulting 2D joint autocorrelation on $(\eta_\Delta,\phi_\Delta)$  compactly represents {\em all} significant correlation structure on 4D angular subspace $(\eta_1,\eta_2,\phi_1,\phi_2)$. 

In Fig.~\ref{Figure2} perspective views are shown of CD joint autocorrelations for four centrality classes of Au-Au collisions at $\sqrt{s_{NN}} =$ 130 GeV. Quantity $\bar N(\hat r - 1)${~\cite{nbar}} represents {\em per-particle} correlations ({\em i.e.,} distribution of average numbers of correlated pairs per final-state particle) and is $O(1)$ for all centralities. Distributions in Fig.~\ref{Figure2} are dominated by a 2D negative peak which is broader and elliptical for peripheral collisions (d) with major axis along $\phi_\Delta$, transitioning smoothly to a narrower and deeper peak symmetric on $(\eta_\Delta,\phi_\Delta)$ for central collisions (a). The negative peak means that unlike-sign charge pairs are more probable than like-sign pairs for small angular separations on pseudorapidity and azimuth, consistent with local charge conservation (suppression of net-charge fluctuations). The vertical axis limits common to all panels were chosen to enhance the visibility of structure at large angular separations as opposed to showing the full depth of the negative peak at $\phi_\Delta = \eta_\Delta = 0$. Note that no CD (charge-dependent) component of elliptic flow is observed at the sensitivity level of these data. 1D projections of Fig.~\ref{Figure2} distributions and their 2D model fits (discussed below) onto individual difference variables $\phi_{\Delta}$ and $\eta_{\Delta}$ are shown in Fig.~\ref{Figure3}. Solid dots and curves (open triangles and dashed curves) correspond to $\eta_{\Delta}$ $(\phi_{\Delta})$ projections. The projections are over the pair acceptances apparent in Fig.~\ref{Figure2}. 

\section{Errors}

Statistical errors for $\hat r$ in Fig.~\ref{Figure1} (central collisions) are $\pm$0.00015 for all bins. Statistical errors for 1D autocorrelations 
are uniform on $\phi_{\Delta}$ (since $\phi$ is a periodic variable) but approximately double as $|\eta_{\Delta}|$ increases from 0 to 2 (due to finite $\eta$ acceptance).  Statistical errors at $\eta_{\Delta} \sim 0$ vary from $\pm$0.00015 for central collisions to $\pm$0.0007 for peripheral collisions, again reflecting the $1/\bar N$ dilution factor. In contrast, statistical errors for $\bar N(\hat r - 1)$ in Fig.~\ref{Figure2} are approximately $\pm 0.2$ (one tick) for $\eta_\Delta \sim 0$ and are independent of centrality. Statistical errors for projections in Fig.~\ref{Figure3} are shown explicitly in that figure by error bars. Systematic errors were estimated as in \cite{meanptprl}. 
Systematic uncertainties associated with two-track inefficiency corrections
and small momentum scale correlation cuts are negligible for this analysis. 
Systematic error due to non-primary backgrounds (dominant source)~\cite{spectra}, whose correlation with true primary particles is unknown, is estimated to be
at most $\pm$7\%, assumed uniform for all $(\eta_{\Delta},\phi_{\Delta})$ in the STAR acceptance.  Contributions from resonance $(\rho^0 , \omega )$ decays are estimated to be at most about 10\% of the negative peaks at $\phi_\Delta = \eta_\Delta = 0$
in Fig.~\ref{Figure2} in the range $|\eta_{\Delta}| < 0.5$, $|\phi_{\Delta}| < 2$~\cite{mevsim}.

\section{Model Fits}

The distributions in Fig.~\ref{Figure2} and their counterpart for p-p collisions~\cite{jeffpp} reveal two asymptotic forms at the centrality limits: a 1D gaussian on $\eta_\Delta$ ({\em uniform} on $\phi_\Delta$) for p-p collisions and a 2D exponential on $(\eta_\Delta,\phi_\Delta)$ for central Au-Au collisions. The two forms may be limiting cases of a single evolving structure, or they may correspond to two independent correlation mechanisms with complementary centrality trends. A preliminary fitting exercise indicated that these 130 GeV Au-Au data do not have sufficient statistical power or centrality range to explore the possibility of a single evolving peak structure. We therefore used the simpler superposition model.

The distributions in Fig.~\ref{Figure2} were fitted with a five-parameter model function 
consisting of a 2D exponential function peaked on both $\eta_{\Delta}$ and $\phi_{\Delta}$ and a 1D gaussian on $\eta_{\Delta}$, constant on $\phi_\Delta$ (the latter motivated by the p-p limiting case~\cite{isrpp,jeffpp}) plus a constant offset, all defined relative to quantity $\hat r - 1$ as 
\bea \label{Eq3}
F & = & A_0 + A_1 \exp \left\{ -\left[ \left(
\frac{\phi_{\Delta}}{ \sigma_{\phi_{\Delta}}} \right)^2  +   \left(
\frac{\eta_{\Delta}}{ \sigma_{\eta_{\Delta}}} \right)^2 \right]
^{\frac{1}{2}} \right\} \nonumber \\
 & + & A_2 \exp \left\{-\left( \frac{\eta_{\Delta}}{1.5 \sqrt{2}} \right)^2 \right\} ~.
\eea
$F$ interpolates between the 1D gaussian peak observed in p-p and the 2D exponential peak observed in central Au-Au collisions. Correlations between amplitudes $A_1$ and $A_2$ were negligible because of the distinct one- and two-dimensional peak shapes. Parameters $\sigma_{\phi_\Delta}$ and $\sigma_{\eta_\Delta}$ are the r.m.s. widths of  the 2D exponential peak when projected onto the respective difference variables.

Best-fit values for varied parameters and $\chi^2/$DoF for the four centralities are listed in Table~\ref{TableI}. The width of the 1D gaussian, most evident near $|\phi_\Delta| \sim \pi$ in Fig.~2(d), was best determined by those peripheral data to be $1.5\pm0.25$ and was held fixed at that value for the other centralities to obtain the amplitude estimates. The observed peripheral Au-Au $\phi_\Delta$ width is definitely larger than the corresponding width for p-p collisions.  Also included is tracking efficiency-correction factor ${\tilde S}$~\cite{extrap}. Total systematic error for efficiency-corrected amplitudes in Table~\ref{TableI} was 11\% (errors added in quadrature).
The model fits indicate that with increasing centrality the 2D exponential peak exhibits 1) strong amplitude increase, 2) significant width reduction and 3) approach to approximately equal widths on $\phi_{\Delta}$ and $\eta_{\Delta}$ for central collisions ({\em cf.} Fig.~\ref{Figure3}; {\em e.g.,} at mid-rapidity $\sigma_{\eta_\Delta} = 0.6$ corresponds to polar angle difference $0.57$, which is directly comparable to $\sigma_{\phi_\Delta}$).

\begin{table}[h]
\caption{\label{TableI}
Parameters and fitting errors (only) for model fits [Eq.~(\ref{Eq3})] to
joint autocorrelation data in Fig.~\ref{Figure2} for centrality bins (a) - (d) (central - peripheral). 
Total systematic error for tracking efficiency-corrected amplitudes is 11\%~\cite{extrap}.
}
\begin{tabular}{|c|c|c|c|c|c|} \hline
centrality  & (d) & (c)  & (b)  & (a)  & error\footnote{Range of fitting errors in percent, from peripheral to central.}(\%) \\
\hline
${\tilde S}$~\cite{extrap} &   1.19    &   1.22    &  1.25    &   1.27   & 8 (syst.)
\\
$\bar{N}$  &   115.5   &   424.9   &  789.3   &   983.0  &  \\
\hline \vspace{-.12in} & & & & & \\
${\tilde S} \bar{N} A_0$     & 0.98  &  0.80  & 0.91  &  0.79  &  11-12 \\
\hline \vspace{-.12in} & & & & & \\
${\tilde S} \bar{N} A_1$     & -4.1  &  -6.8  &  -7.7  &  -7.7  &  6-4  \\
$\sigma_{\phi_{\Delta}}$ & 0.94  &  0.75  &  0.72   &  0.72  &  11-5   \\
$\sigma_{\eta_{\Delta}}$ & 0.66  &  0.59  &  0.58   &  0.58   & 10-5  \\
\hline \vspace{-.12in} & & & & & \\
${\tilde S} \bar{N} A_2$     & -0.51 & -0.11  &  -0.15 & -0.021  &
0.17-0.19
\footnote{Magnitude of fitting errors.} \\
\hline \vspace{-.12in} & & & & & \\
$\chi^2$/DoF & $\frac{380}{315}$ & $\frac{315}{315}$ & $\frac{314}{315}$ & $\frac{329}{315}$ & \\  
\hline 
\end{tabular}
\end{table}

\section{Discussion}

This analysis demonstrates for the first time that charge-dependent angular correlations for central Au-Au collisions differ dramatically from those for p-p collisions. CD angular correlations for p-p collisions are dominated by a 1D negative gaussian peak on $\eta_{\Delta}$ with $\sigma_{\eta_\Delta} \simeq 1$~\cite{isrpp,jeffpp}, conventionally associated with longitudinal charge ordering on $z$ during string fragmentation~\cite{lund}, plus a 2D gaussian peak associated with quantum correlations.
For the most peripheral Au-Au centrality (d) in this analysis we observe CD correlation structure intermediate between p-p and central Au-Au collisions, consistent with the fact that collision events in centrality class (d) for these 130 GeV data are not very peripheral: they contain about 100 particles in the STAR acceptance (see Table~\ref{TableI}). In central Au-Au collisions the 1D gaussian peak is no longer detectable. Instead, a large-amplitude 2D negative exponential peak dominates the correlation structure, with similar widths on $\eta_\Delta$ and $\phi_\Delta$ much reduced from those measured in p-p collisions.

\begin{figure}[h]
\includegraphics[keepaspectratio,width=3.3in]{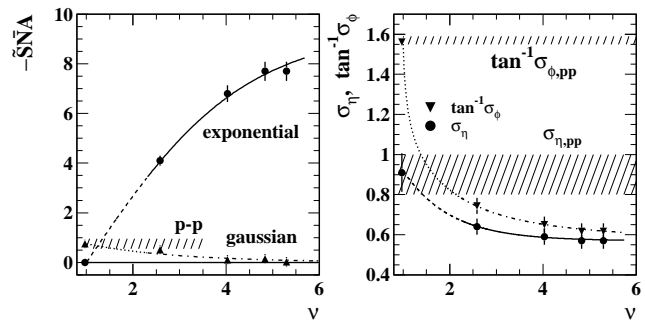}
\caption{\label{Figure4} Left panel: Efficiency corrected correlation amplitudes for 2D exponential (dots) and 1D gaussian (triangles) components from Table~\ref{TableI} for negative peaks in Fig.~\ref{Figure2} are plotted on mean path length $\nu$~\cite{nu}.
Right panel: Fitted widths $\sigma_{\eta_\Delta}$ (dots) and $\tan^{-1}\, \sigma_{\phi_\Delta}$ (triangles) are plotted on $\nu$.  Plotting variable $\tan^{-1}$ permits the divergent p-p $\sigma_{\phi_\Delta}$ value to be included. Hatched regions and $\nu = 1$ data points summarize p-p limiting values. Curves guide the eye.}
\end{figure}

Variations of peak amplitudes and widths with Au-Au centrality are shown in Fig.~\ref{Figure4}, along with p-p limiting cases (cross-hatched bands) from STAR p-p data at 200 GeV~\cite{jeffpp}, consistent with ISR p-p data at 52.5 GeV~\cite{isrpp}. The p-p data points in Fig.~\ref{Figure4} (values at $\nu = 1$) indicate the amplitude and r.m.s. width of the 1D gaussian on $\eta_\Delta$, the uniformity of that correlation on $\phi_\Delta$ ($\sigma_{\phi_\Delta} \gg 1$) and the absence of a 2D exponential on $(\eta_\Delta,\phi_\Delta)$ in the fit residuals, represented by the solid dot in the left panel at $\nu = 1$. Comparison of the low-$p_t$ ($0.15 \leq p_t \leq 0.5$ GeV/c) p-p results with the present Au-Au results is qualitative but reasonable given the similarity in shape of the Au-Au CD correlations for $0.15 \leq p_t \leq 0.5$ (discussed below) to those in Fig.~\ref{Figure2}. 

The collision centrality is represented by mean participant path length $\nu$~\cite{nu}, defined as the average number of nucleons encountered by a participant nucleon. That centrality measure is desirable because it permits comparisons with p-A collisions, initial-state scattering should follow a trend linear in $\nu$ and $\nu$ also provides an estimate (proportionality) of final-state pathlength. {

We adopt the strategy of plotting $\tan^{-1}(\sigma_{\phi_\Delta})$ rather than $\sigma_{\phi_\Delta}$ so as to include the p-p `infinite azimuth width' on the same plot, since that distribution is approximately uniform on $\phi$. Interpolations among the measured Au-Au points are sketched by the solid and dash-dot curves. {\em Extrapolations} to corresponding p-p values are sketched by the dashed and dotted curves. The extrapolations contain {\em substantial uncertainties} in relating p-p to mid-peripheral Au-Au results. \em Efficiency-corrected} per-particle correlation amplitudes $\tilde S\bar NA$ for central Au-Au collisions exceed in magnitude those for p-p collisions {\em by a factor 10}. The dramatic shape and amplitude changes strongly contradict a p-p linear superposition hypothesis~\cite{nbar} for all but the most peripheral Au-Au collisions.

These results for net-charge angular correlations suggest that CD correlations in Au-Au collisions, as in p-p collisions, derive from configuration-space charge ordering as a consequence of local charge conservation during hadronization, but the hadronization geometry changes from 1D ($\eta$) in p-p collisions to {\em at least} 2D ($\eta,\phi$) in central Au-Au collisions, leading to an approach to angular symmetry on $(\eta_\Delta,\phi_\Delta)$. Transverse charge ordering (on $p_t$) is also possible but is studied in a separate analysis. Hadronic rescattering in A-A collisions could reduce the CD correlation amplitude at large $\phi_\Delta$ but would also reduce the width on $\eta_\Delta$ and therefore cannot be solely responsible for the nearly symmetric peak shape in central Au-Au collisions.  In Fig.~\ref{Figure4} the contribution from 1D charge ordering (gaussian peak on $\eta_\Delta$) is already substantially reduced for centrality (d) ($\nu \sim 2.5$) in favor of the symmetric component (exponential peak). 

A hadron-opaque medium in more central collisions may contribute to the newly-observed {\em exponential} peak shape. An exponential distribution on pair opening angle [radius on $(\eta,\phi)$] is consistent with: 1) correlations detected only if both members of a correlated pair are not significantly scattered, 2) scattering probability determined by a mean free path, 3) mean path length in the medium increasing monotonically with pair opening angle. That rescattering picture assumes that CD correlations do not result from hadronization outside the medium. Contributions from charge ordering in jet fragmentation were studied by splitting central Au-Au data at $p_t = 0.5$ GeV/$c$, below which jet fragments should be negligible. Negative peak structures as in Fig.~\ref{Figure2} were observed to dominate both subsamples, although the amplitudes were not identical. 






{\sc hijing}~\cite{jetquench} and {\sc rqmd}~\cite{rqmd,starhbt} charge-dependent angular correlations qualitatively disagree with data. {\sc hijing} charge-dependent correlations are determined by the Lund model~\cite{lund} {\em via} {\sc pythia}~\cite{pythia}, and are consequently consistent with p-p 1D string fragmentation for all A-A centralities: a 1D gaussian on $\eta_\Delta$ with amplitude about 10\% of the exponential peak in Fig.~\ref{Figure2} (a). RQMD, representing mainly resonance decays and hadronic rescattering, exhibits a broad 2D gaussian on ($\eta_\Delta,\phi_\Delta $), with amplitude also about 10\% of the exponential peak in the data for central collisions. 
Large-scale correlations as in Fig.~\ref{Figure1} observed for US {\em and}\, LS pairs in data are consistent with local charge ordering but {\em inconsistent} with CD correlations from decays of hadronic resonances such as the $\rho^0$, which affect only the US pair type. That observation further argues against a resonance-gas scenario. 

Measurements of net-charge fluctuations have been advocated as a probe of heavy-ion collisions. Predictions of dramatic suppression of net-charge fluctuations in the case of QGP formation based on entropy arguments~\cite{jeon} refer by implication to an {\em integral} of net-charge angular correlations over a detector acceptance. Phenix observed net-charge fluctuations in Au-Au at 130 GeV~\cite{phen-net} slightly reduced from `stochastic behavior' and independent of collision centrality. The data were consistent with RQMD representing a resonance gas. STAR observed net-charge fluctuations in Au-Au at 200 GeV~\cite{star-net} intermediate between what is expected from canonical suppression in a partial acceptance and a resonance gas, again with little or no centrality dependence. Those conclusions are in sharp contrast to what we observe in the present analysis.

It is important to note that net-charge fluctuations within a given detector acceptance integrate CD joint autocorrelations such as those presented in this paper (within a constant offset) over that acceptance, as described in~\cite{inverse}. As integral quantities, fluctuation measurements are insensitive to the {\em differential structure} of angular correlations. In the present analysis we observe dramatic changes in differential structure (10-fold amplitude increase, nearly two-fold width reduction) while corresponding peak integrals  exhibit only modest change with collision centrality (integrals of observed CD peaks using peak parameters in Table~\ref{TableI} increase linearly in magnitude on $\nu$ by about 20\%). We suggest that the theoretical connection between net-charge fluctuation suppression and QGP formation, currently based only on large-scale integral measures, should be re-examined in the more differential context of CD autocorrelation structure.

\section{Summary}

In summary, we have measured charge-dependent angular correlations on pseudorapidity and azimuth difference variables $(\eta_1 - \eta_2)$ and $(\phi_1 - \phi_2)$ for Au-Au collisions at $\sqrt{s_{NN}}$ = 130 GeV. The data are consistent with {\em local} charge conservation or canonical suppression of net-charge fluctuations, evolving from 1D (along $\eta$) color-string fragmentation in p-p collisions to exponentially-attenuated (on opening angle) 2D charge-ordered emission from a hadron-opaque medium in central Au-Au collisions. 
The transition from 1D to 2D correlation structure occurs rapidly with increasing collision centrality.  These results are qualitatively inconsistent with predictions from standard Monte Carlo collision models typically applied to single-particle differential distributions and integrated yields from relativistic heavy-ion collisions.
Charge-dependent angular autocorrelations provide unique {\em differential} access to the changing geometry of hadronization and hadronic rescattering as the energy density and spatial extent of A-A collisions vary with centrality. 


We thank the RHIC Operations Group and RCF at BNL, and the
NERSC Center at LBNL for their support. This work was supported
in part by the HENP Divisions of the Office of Science of the U.S.
DOE; the U.S. NSF; the BMBF of Germany; IN2P3, RA, RPL, and
EMN of France; EPSRC of the United Kingdom; FAPESP of Brazil;
the Russian Ministry of Science and Technology; the Ministry of
Education and the NNSFC of China; IRP and GA of the Czech Republic,
FOM of the Netherlands, DAE, DST, and CSIR of the Government
of India; Swiss NSF; the Polish State Committee for Scientific 
Research; STAA of Slovakia, and the Korea Sci. \& Eng. Foundation.


\begin{thebibliography}{9}
\bibitem{average} Averaging rather than simple projection is an essential feature of autocorrelations required to properly account for acceptance effects in two-particle spaces.
\bibitem{nu} Centrality measure $\nu$ estimates the mean participant path length as a number of encountered nucleons. For this analysis $\nu \equiv 5.5 \, (N/N_0)^{1/3} \simeq 5.5\, (N_{part}/N_{part,max})^{1/3} \simeq 2N_{bin}/N_{part}$, based on Glauber-model simulations. $N_{part}$ is the number of participants, $N_{bin}$ the number of binary collisions, and $N_0$ is the upper half-maximum endpoint of the minimum-bias data distribution plotted as $d\sigma/dN^{1/4}_{}$.
\bibitem{extrap} Extrapolation factors ${\tilde S}$ for $\bar{N}(\hat r - 1 )$
provide corrections to amplitudes $A_0$, $A_1$ and $A_2$ for background contamination and tracking inefficiency~\cite{spectra}. Systematic error in ${\tilde S}$ was estimated to be $\pm$8\%.
\end{thebibliography}
\end{document}